\documentclass[conference,a4paper]{IEEEtran}

\usepackage{listings}
\renewcommand{\lstlistingname}{\bfseries Listing}
\makeatletter
\def\fnum@lstlisting{%
  \lstlistingname
  \ifx\lst@@caption\@empty\else~\thelstlisting\normalfont\fi}%
\makeatother

\usepackage{cite}

\ifCLASSINFOpdf
  	\usepackage[pdftex]{graphicx}
\else
\fi
\begin{document}
%
\title{CppSs -- a C++ Library for Efficient Task Parallelism}

\author{\IEEEauthorblockN{Steffen Brinkmann and Jos\'e Gracia}
\IEEEauthorblockA{High Performance Computing Centre Stuttgart (HLRS)\\
University of Stuttgart\\
70550 Stuttgart, Germany\\
E-mail: \{brinkmann,gracia\}@hlrs.de}

}


%


\maketitle

\begin{abstract}
We present the C++ library CppSs (C++ super-scalar), which 
provides efficient task-parallelism
without the need for special compilers or other software. Any C++ compiler
that supports C++11 is sufficient.
CppSs features different directionality clauses for defining data dependencies. While
the variable argument lists of the taskified functions are evaluated
at compile time, the resulting task dependencies are fixed by the runtime value
of the arguments and are thus analysed at runtime. With CppSs, we provide
task-parallelism using merely native C++.
\end{abstract}



%
\IEEEpeerreviewmaketitle

\section{Introduction}

Programming models implementing task-parallelism play a major role when preparing
code for modern architectures with many cores per node and thousands of nodes
per cluster. 
In high performance computing, a common approach for achieving the best 
parallel performance is to apply the message passing interface (MPI)\cite{mpi-web}
for inter-node communication and a shared-memory
programming model for intra-node parallelisation. This way, the communication 
overhead of pure MPI applications can be overcome.

Shared memory models are also crucial when using single node computers as
there are systems consisting of hundreds or even thousands of processing 
units accessing the same memory address space. These systems offer great
parallelism to the developer. But utilising the processing units evenly,
so that they can run efficiently, is a non-trivial task.

Many scientific applications are based on processing large amounts 
of data. Usually, the processing of this data can be split up and some of these 
chunks have to be executed in a well defined order while others are
independent. This is the level on which task based programming models are
employed. We will call the chunks of work to be processed tasks, while the 
appearances in the code (e.g., if they are implemented as functions, methods or 
subroutines) are going to be called task instances.

\begin{figure}
(a)\includegraphics[width=.20\textwidth]{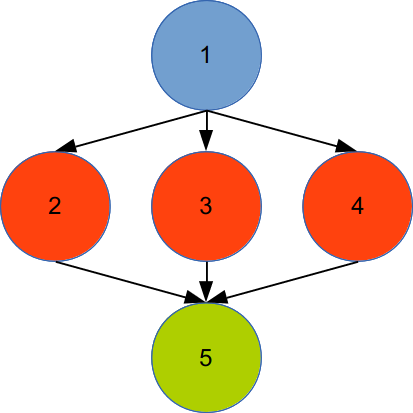}
(b)\includegraphics[width=.22\textwidth]{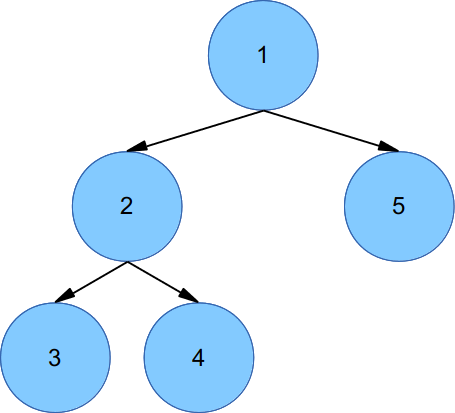}
\caption{(a) Example of fork-join-parallelism. After task 1 the execution 
thread is forked. \mbox{(b) Example} of nested parallelism. Task 1 spawns tasks 2 and 5. Before task 5 
is created, task 3 and 4 are spawned, hence the numbering.}
\label{fig:forkjoin}
\end{figure}

The dependencies between tasks can be stated explicitly by the programmer or
inferred automatically by some kind of preprocessing of the code. In the 
case of fork-join-models (e.g. OpenMP\cite{openmp-web}), all tasks after a
``fork'' are (potentially) parallel while code after the ``join'' and all consecutive
forks depend on them. For example, in figure~\ref{fig:forkjoin}a),  tasks 2, 
3 and 4 can run in parallel, if sufficient processing units are available. 
Task 5 cannot be executed before all other tasks have finished. In programming
models which support nesting (e.g. Cilk\cite{cilk1}), the dependencies 
can sometimes be derived from the placement of the calls (see 
Figure~\ref{fig:forkjoin}b)).

In many implementations of task based programming models, the data
dependencies are specified explicitly by the programmer (e.g. 
SMPSs\cite{text-web}, OMPSs\cite{ompss-web}, StarPU\cite{starpu-1} and
XKAAPI\cite{xkaapi-web}). This allows for more complex dependency graphs
and therefore more possibilities to adjust the parallelisation to
the code, the amount of data and the architecture. However, these 
implementations suffer from a number of disadvantages:
\begin{itemize}
\item The tasks and/or task instances and their dependencies have to be
	marked by special directives, usually within a \verb|#pragma| in C or
	using special comments in Fortran. These use keywords and syntax 
	which is not part of the actual language and which
	the programmer needs to learn.
\item In order to compile the instrumented code, the programmer needs a
	special compiler or preprocessor. She depends on this additional
	software to be available on the desired platform, which is not generally
	the case. 
\item The need for special compilers also poses additional work to 
	system administrators who will be asked by the programmer to install 
	the specific compiler used in the application.
\item The code of the programming model implementation itself becomes
	more difficult to maintain and usually at least one additional compile step is
	introduced when compiling the user code.
\end{itemize}

In order to avoid these inconveniences, we developed a pure C/C++ library,
which allows functions to be marked as tasks and to execute them 
asynchronously. The programmer still needs to prepare the code looking for
the parts feasible for parallelisation and separate them into functions. Also,
it is still necessary to instrument the code with the CppSs API.
But contrary to the implementations mentioned above, this is achieved using 
standard C++11 syntax instead of an ``imposed'' pragma language.

To execute the application serially, e.g. for debugging, the programmer can 
define the macro \texttt{NO\_CPPSS},
which bypasses the creation of additional threads and converts the tasks instances
into normal function calls.

In the following, we will 
illustrate the usage (Section~\ref{sec:usage})
and present the basic implementation of the library CppSs (Section~\ref{sec:impl}).
Lastly, we will sum up our conclusions in Section~\ref{sec:concl}.

\section{CppSs - usage}
\label{sec:usage}

CppSs is a library which compiles on any system with a working C++ compiler.
The C++11 features necessary for CppSs are provided by the GNU 
compiler of version 4.6 or higher and the Intel compiler of version 13 or higher.

In order to use CppSs, the programmer only needs to include the header 
\texttt{CppSs.h} and link against the library \texttt{libcppss.so}.
All of CppSs' application programming interface (API) functions are
declared in the namespace {\tt CppSs} to
avoid overlap with other libraries' functions. In the following,
the CppSs API is introduced presenting the declaration of tasks
(Section~\ref{sec:decl}), the initialisation and finishing of the
parallel execution (Section~\ref{sec:initfinish}) and setting barriers
(Section~\ref{sec:barriers}). Finally, we will give a minimal
example putting everything together in Section~\ref{sec:example}.

\subsection{Declaring Tasks}
\label{sec:decl}

Parallelisation with CppSs relies on functions with well defined 
directionality of their parameters. Loop parallelisation and anonymous
code blocks are not supported. 

To convert a function into a task,
the programmer has to call the API function {\tt MakeTask}, which takes the
following parameters (see listing in Figure ~\ref{lst:minimal}):

\begin{itemize}
\item a pointer to the function,
\item an initialiser list containing directionality specifiers for each 
	function parameter,
\item (optional) a string with the function name for debugging purposes and
\item (optional) a priority level, which is ignored in the present version.
	Future versions will provide one or more priority queues.
\end{itemize}

\begin{figure}
\begin{lstlisting}
void func1(int *a1, double *a2, double *b)
{
	//...
}
auto func1_task = CppSs::MakeTask(func1,
                                  {INOUT,IN,OUT},
                                  "func1");
\end{lstlisting}
\caption[Defining and taskifying a function]{
	Defining and taskifying a function. The return value is a functor, 
	i.e., an object which overloads
	the parenthesis operator. Hence it can be ``called'' like a function.}
\label{lst:minimal}
\end{figure}

It is required that the arguments of the taskified function
which are intended to cause dependencies are pointers.
These can be used to access arrays, built-in types or any other data
structure. However, potential overlap with
other data structures is not detected. The directionality specifier must be one of
{\tt IN, OUT, INOUT}, {\tt REDUCTION} or {\tt PARAMETER}.
The latter is used for arguments which are not to be interpreted as a 
potential dependency and must be of a built-in numerical type.
The effect of each of the directionality specifiers are described in 
the following:

\paragraph{IN} The task treats this argument as input. 
It will not be executed until all task 
instantiations which were called before the function and which 
write to this argument (i.e. have an {\tt OUT, INOUT} or {\tt REDUCTION}
specifier for the same argument value) have finished.

\paragraph{OUT} The task treats this argument as output. The content
of the variable or array pointed to is (possibly) overwritten. This
affects functions with an {\tt IN} or {\tt INOUT} specifier for the
same argument value.

\paragraph{INOUT} The task intends to read from and write to this 
argument value. It will be dependent on the last task writing to this
memory address. The following tasks reading from this memory address
will be dependent on this task.

\paragraph{REDUCTION} Similar to {\tt INOUT}. The task intends to read 
from and write to this argument value. In contrast to {\tt INOUT}, 
the tasks with a {\tt REDUCTION} clause will depend on other tasks
with a {\tt REDUCTION} clause on the same argument value.

\paragraph{PARAMETER} The argument is treated as a parameter. It will
be ignored for the dependency analysis.

The return type of {\tt MakeTask} is an internal template type, which
includes the argument types of the taskified function, thus we
recommend to use the C++11 keyword {\tt auto}. 

For convenience two macros were defined that wrap the call to {\tt MakeTask}.
The three calls in Figure~\ref{lst:macros} are equivalent.

\begin{figure}
\begin{lstlisting}
auto func_task = CppSs::MakeTask(func,
                                 {INOUT,IN,OUT},
                                 "func");
auto func_task = CPPSS_TASK(func,
                            {INOUT,IN,OUT});
CPPSS_TASKIFY(func,{INOUT,IN,OUT})
\end{lstlisting}

\caption[Convenience macros for task declaration]{
	Convenience macros for task declaration. These three lines translate into
	the same binary code. Hence only one of them should be used.}
\label{lst:macros}
\end{figure}

\subsection{Init and Finish}
\label{sec:initfinish}

The next instrumentation to be inserted in the application code
is calls to {\tt Init} and {\tt Finish}.
These calls must be called before and after each task, respectively.
While {\tt Finish} takes no arguments, {\tt Init} takes two optional arguments, 
namely

\begin{itemize}
\item the number of threads and
\item the reporting level.
\end{itemize}

The number of threads must be any positive integer. If none is given, the default
is 2. The reporting level must be one of {\tt ERROR, WARNING, INFO} or {\tt DEBUG}, 
which causes increasing amount of output. The default is {\tt WARNING}.

{\tt Init} will instantiate a runtime system which enables the queuing and 
asynchronous execution of tasks. The runtime will create one thread less 
than the number of threads specified in the call to \texttt{Init} as 
the main thread will also execute tasks. The threads will be constructed
using the standard library {\tt std::thread} class. This way portability
is granted for each system which provides a C++11 compiler.

{\tt Finish} will wait for all the tasks to be finished and destruct all threads,
queues and the runtime.

\subsection{Barriers}
\label{sec:barriers}

With the API function {\tt Barrier} it is possible to halt the main execution
thread, i.e. the code outside of tasks, until all tasks instantiated
so far have finished. The call takes no arguments. The call to {\tt Finish}
contains a call to {\tt Barrier}.

\subsection{Minimal example}
\label{sec:example}

\begin{figure}
\centering
\includegraphics[width=0.2\textwidth]{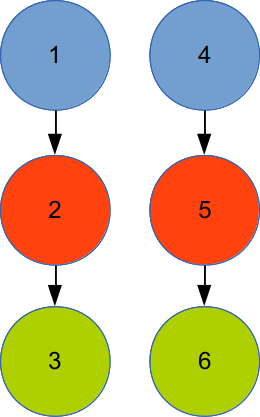}
\caption{Task dependency graph of the minimal example in listing in Figure ~\ref{lst:example}.
The blue nodes (1 and 4) represent task function {\tt set\_task}, the red nodes 
(2 and 5) {\tt increment\_task} and the green nodes (3 and 6) {\tt output\_task}. }
\label{fig:example}
\end{figure}

To sum up the API usage, we compile everything into a small example, shown in 
Figure~\ref{lst:example}. Internally, it produces the dependency graph shown in 
Figure~\ref{fig:example} and prints the output shown in Figure~\ref{lst:output}.

\begin{figure}
\begin{lstlisting}
#include <iostream>
#include "CppSs.h"

#define N_THREADS 2

void set(int *a, int b)
{
    (*a) = b;
}
CPPSS_TASKIFY(set,{OUT,PARAMETER})

void increment(int *a)
{
    ++(*a);
}
CPPSS_TASKIFY(increment,{INOUT})

void output(int *a)
{
    std::cout << (*a) << std::endl;
}
CPPSS_TASKIFY(output,{IN})

int main(void)
{
    int a[] = {1,11};
	
    CppSs::Init(N_THREADS,INFO);
    
    for (unsigned i=0; i < 2; ++i){
        set_task(&(a[i]), i);
    	increment_task(&(a[0]));
    	output_task(&(a[0]));
    }
    
    CppSs::Finish();

    return 0;
}
\end{lstlisting}

\caption[Minimal complete example for CppSs]{
	Minimal complete example for CppSs. This code will produce a dependency
	graph as shown in Figure~\ref{fig:example}. The output will be similar to
	listing in Figure~\ref{lst:output}.}
\label{lst:example}
\end{figure}

\begin{figure}
\begin{lstlisting}[
language=
]
- 13:32:45.207 INFO:  ### CppSs::Init ###
- 13:32:45.207 INFO: adding worker: 1 of 2
- 13:32:45.207 INFO: Running on 2 threads.
1
2
- 13:32:45.207 INFO: Executed 6 tasks.
- 13:32:45.207 INFO:  ### CppSs::Finish ###
\end{lstlisting}
\caption[Output from minimal example]{
	Output from minimal example from listing in Figure~\ref{lst:example}.}
\label{lst:output}
\end{figure}

\section{CppSs - implementation with variadic templates}
\label{sec:impl}

The major design paradigm for CppSs was to avoid usage of external libraries.
All code should be compilable with a standard C++ compiler. In order to
achieve this goal, several features of C++11 were used, the most prominent
one being variadic templates\cite{gregor08:VariadicTemplates}. These are
of central importance as the objects representing a task and
an instance of a task are implemented as variadic templates, the function
arguments of the taskified function being the template arguments.
This is necessary because a function which the application programmer
wants to taskify can have any number and type of arguments.
These arguments are known at compile time, so an implementation with 
variadic templates is the most efficient way to handle variable argument
lists.

An excerpt of the {\tt Task\_functor} class declaration which stores
the taskified function is shown in Figure~\ref{lst:taskfunctor}.

\begin{figure}
\begin{lstlisting}
template<typename... ARGS>
class Task_functor : public Task_functor_base
{
    //...
    void (*m_f) (ARGS...);
}
\end{lstlisting}
\caption[Excerpt from the class declaration of {\tt Task\_functor}]{
	Excerpt from the class declaration of {\tt Task\_functor} which stores
	the taskified function. The member declaration shows the pointer to
	the actual function with a variable argument list.}
\label{lst:taskfunctor}
\end{figure}

\begin{figure}

\begin{lstlisting}[
	basicstyle=\scriptsize\ttfamily
]
template <typename fun, size_t i>
struct get_types_helper {
    static void get_types(
       std::vector<std::type_info const*> &types) {
        get_types_helper<fun, i-1>::get_types(types);
        types.push_back(&typeid(typename 
           function_traits<fun>::template arg<i-1>::type));
    }
};

template <typename fun>
struct get_types_helper<fun,0> {
    static void get_types(
       std::vector<std::type_info const*> &types) {}
};

template <typename fun>
void get_types(std::vector<std::type_info const*> &types) {
    get_types_helper<fun, function_traits<fun>::nargs>::\
       get_types(types);
}
\end{lstlisting}

\caption[Template functions to process argument types at compile time]{
	Template functions to process argument types at compile time. A call to
	{\tt get\_types<function>(types)} will recursively get the type of each of
	{\tt function}'s arguments and place their type in the array {\tt types}.}
\label{lst:gettypes}
\end{figure}

In order to process the variable argument list at compile time,
recursive template evaluation is necessary. For instance, 
the set of template functions used to retrieve the types of 
the task function arguments is shown in Figure~\ref{lst:gettypes}.

\section{Conclusion}
\label{sec:concl}

We developed a pure C/C++ library, which allows functions in C/C++ source code
to be marked as tasks, specify their dependencies and to execute them 
asynchronously. Contrary to other similar task based programming models
like OpenMP, SMPSs or OMPSs, 
no preprocessor directives are necessary and the instrumented
code will compile with any compiler, which supports C++11 features such 
as variadic templates, smart pointers and initializer lists. 
The smallest versions that qualify 
of the GNU compiler collection (gcc) and the Intel C compiler (icc), both of 
which are widely available, are gcc 4.6 and icc 13.

The current version is capable of constructing the task dependency graph
and execute the tasks asynchronously. Several directionality clauses are 
available.

The code was checked for correctness but has still to prove scalability in
realistic scenarios. First performance tests showed more than three times faster
execution when running on four cores compared with the serial
version of the same algorithm. We believe that these results can be enhanced
by revising the implementation of the queueing and dequeueing as well as
the creation and destruction of task functor instances.


\section*{Acknowledgment}

The authors acknowledge support by the H4H
project funded by the German Federal Ministry for Education and Research
(grant number 01IS10036B) within the ITEA2 framework (grant number 09011).




\bibliographystyle{IEEEtran}

\bibliography{bibliography}

\begin{thebibliography}{1}
\providecommand{\url}[1]{#1}
\csname url@samestyle\endcsname
\providecommand{\newblock}{\relax}
\providecommand{\bibinfo}[2]{#2}
\providecommand{\BIBentrySTDinterwordspacing}{\spaceskip=0pt\relax}
\providecommand{\BIBentryALTinterwordstretchfactor}{4}
\providecommand{\BIBentryALTinterwordspacing}{\spaceskip=\fontdimen2\font plus
\BIBentryALTinterwordstretchfactor\fontdimen3\font minus
  \fontdimen4\font\relax}
\providecommand{\BIBforeignlanguage}[2]{{%
\expandafter\ifx\csname l@#1\endcsname\relax
\typeout{** WARNING: IEEEtran.bst: No hyphenation pattern has been}%
\typeout{** loaded for the language `#1'. Using the pattern for}%
\typeout{** the default language instead.}%
\else
\language=\csname l@#1\endcsname
\fi
#2}}
\providecommand{\BIBdecl}{\relax}
\BIBdecl

\bibitem{mpi-web}
\BIBentryALTinterwordspacing
{Message Passing Interface Forum} (http://www.mpi-forum.org/) [retrieved: 09,
  2013]. [Online]. Available: \url{http://www.mpi-forum.org/}
\BIBentrySTDinterwordspacing

\bibitem{openmp-web}
\BIBentryALTinterwordspacing
{OpenMP} (http://openmp.org/wp/) [retrieved: 09, 2013]. [Online]. Available:
  \url{http://openmp.org/wp/}
\BIBentrySTDinterwordspacing

\bibitem{cilk1}
\BIBentryALTinterwordspacing
R.~D. Blumofe, C.~F. Joerg, B.~C. Kuszmaul, C.~E. Leiserson, K.~H. Randall, and
  Y.~Zhou, ``Cilk: an efficient multithreaded runtime system,'' in
  \emph{Proceedings of the fifth ACM SIGPLAN symposium on Principles and
  practice of parallel programming}, ser. PPOPP '95.\hskip 1em plus 0.5em minus
  0.4em\relax New York, NY, USA: ACM, 1995, pp. 207--216. [Online]. Available:
  \url{http://doi.acm.org/10.1145/209936.209958}
\BIBentrySTDinterwordspacing

\bibitem{text-web}
\BIBentryALTinterwordspacing
{TEXT - Towards EXaflop applicaTions} (http://www.project-text.eu) [retrieved:
  09, 2013]. [Online]. Available: \url{http://www.project-text.eu/}
\BIBentrySTDinterwordspacing

\bibitem{ompss-web}
\BIBentryALTinterwordspacing
{The OmpSs Programming Model} (http://pm.bsc.es/ompss) [retrieved: 09, 2013].
  [Online]. Available: \url{http://pm.bsc.es/ompss}
\BIBentrySTDinterwordspacing

\bibitem{starpu-1}
C.~Augonnet, S.~Thibault, R.~Namyst, and P.-A. Wacrenier, \emph{{StarPU: A
  Unified Platform for Task Scheduling on Heterogeneous Multicore
  Architectures}}, 2009.

\bibitem{xkaapi-web}
\BIBentryALTinterwordspacing
{XKAAPI - Kernel for Adaptative, Asynchronous Parallel and Interactive
  programming} (http://kaapi.gforge.inria.fr/) [retrieved: 09, 2013]. [Online].
  Available: \url{http://kaapi.gforge.inria.fr/}
\BIBentrySTDinterwordspacing

\bibitem{gregor08:VariadicTemplates}
\BIBentryALTinterwordspacing
D.~Gregor and J.~J{\"a}rvi, ``Variadic templates for {C++0x},'' \emph{Journal
  of Object Technology}, vol.~7, no.~2, p. 31{\textendash}51, 02/2008 2008.
  [Online]. Available:
  \url{http://www.jot.fm/issues/issue\_2008\_02/article2.pdf}
\BIBentrySTDinterwordspacing

\end{thebibliography}

%
%
%


\end{document}